\documentclass[12pt]{iopart}
\usepackage{graphicx}

\def\etal{{\it et\thinspace al.}\ }
\def\eion{{(e~+~ion)}\ }
\def\ra{{$\rightarrow$}\ }
\def\fexvii{{\rm Fe~\sc xvii}\ }
\def\fexviii{{\rm Fe~\sc xviii}\ }

\def\en{{$n$\ }}

\newcommand{\be}{\begin{equation}}
\newcommand{\ee}{\end{equation}}
\begin{document}

\title[R-Matrix calculations for opacities~.I]{R-Matrix calculations for
opacities:~I. Methodology and computations} 

\author{A K Pradhan$^{1,2,3}$, S N Nahar$^1$, W
Eissner$^4$\footnote{Deceased 2022}}

\address{$^1$ Department of Astronomy, $^2$ Biophysics Graduate Program,
$^3$ Chemical Physics Program,
The Ohio State University, Columbus, Ohio 43210, USA}
\address{$^4$ Instit\"ut f\"ur Theoretische Physik, Teilinstitut 1,
 70550 Stuttgart, Germany}
\vspace{10pt}

\begin{abstract}
An extended version of the R-matrix methodology is presented for
calculation of radiative parameters for improved plasma opacities. 
Contrast and comparisons
with existing methods primarily relying on the Distorted Wave (DW)
approximation are discussed to verify accuracy and resolve outstanding issues,
particularly with reference to the Opacity Project (OP). Among the
improvements incorporated are: (i) large-scale Breit-Pauli R-matrix
(BPRM) calculations
for complex atomic systems including fine structure, (ii) convergent
close coupling wave function expansions for the \eion system to
compute oscillator strengths and photoionization cross sections, 
(iii) open and closed shell
iron ions of interest in astrophysics and experiments, (iv) a 
treatment for plasma broadening of autoionizing resonances as function
of energy-temperature-density dependent cross sections, (v) a
"top-up" procedure to compare convergence with R-matrix calculations
for highly excited
levels, and (vi) spectroscopic identification of resonances
and bound \eion levels.
The present R-matrix monochromatic opacity spectra 
are fundamentally different from OP and lead to enhanced
Rosseland and Planck mean opacities.
An outline of the work reported in other papers in this series and
those in progress is presented.
Based on the present re-examination of the OP work, it is
evident that opacities of heavy elements require revisions in high
temperature-density plasma sources.
\end{abstract}

%
%
%
%
%

\section{Introduction}

Opacity is due to interaction of radiation with matter.
It is a fundamental parameter in plasma, astrophysics, and atomic physics that
determines radiation transport,
and entails absorption and scattering of photons by atoms at
all frequencies of radiation prevalent in a given environment. Methods
for calculating opacities are well-established, and essentially
involve
the atomic physics of bound-bound and bound-free transition 
probabilities incorporated within an equation-of-state (EOS) of the plasma.
However, in practice complexities arise owing to several physical factors that
influence the accurate determination of opacity, and are
addressed in this series of papers. As this work is an extension of
the Opacity Project (hereafter OP), 
we first briefly outline OP and its calculations described under the
{\it Atomic data for opacities} (hereafter ADOC) series 
of papers, and their limitations. Next, we describe the extensions and
improvements over OP in
the present series {\it R-Matrix calculations for opacities}
(hereafter RMOP), subsequently referred to as papers RMOP1, RMOP2,
RMOP3, RMOP4.

\subsection{The Opacity Project}

 The OP work by M.J. Seaton and collaborators \cite{adoc1,symp,op} 
and reference
therein] was devoted to the
development of a framework for calculation of opacities based on the
close coupling approximation implemented in the 
powerful R-Matrix (RM) method by P.G. Burke and collaborators, and employed
extensively for accurate calculations of a variety of radiative and
collisional atomic processes \cite{b11,adoc2,aas}.
The OP work entailed an EOS for stellar interior plasmas
based on the "chemical picture" by D. Mihalas, D.G. Hummer and W. D\"appen
(named MHD-EOS \cite{mhd}), that connects physically with OP
atomic data via an {\it occupation probability} factor of
ionization fractions, level populations, and partition function in the
modified Saha-Boltzmann equations that accounts for plasma interactions.

 Despite unprecedented 
effort and advances, the OP R-matrix work reported in ADOC faced several then
intractable difficulties that limited the scope of atomic calculations.
Primarily, the limitations were due to computational constraints which,
in turn, did not enable accounting for important physical effects and a 
complete
R-matrix calculation of atomic opacities. The main features and
deficiencies of OP are as follows: (I) The calculations were in LS
coupling neglecting relativistic fine structure, (II) The close coupling
(hereafter CC) 
wavefunction expansion for the target or the core ion in the \eion 
system included 
only a few ground configuration LS terms, (III) Inner-shell excitations
could not be included owing to the restricted target ion expansion, (IV)
While autoionizing resonances in bound-free photoionization cross sections were
delineated within the few excited target terms,
(V) Total angular and spin \eion
symmetries with large orbital angular-spin quantum numbers were not computed.
All of these factors are crucial for a
complete and accurate opacity calculation.
 Therefore, the OP work incorporated a relatively small subset of R-matrix data.
Rather, most of the opacities contributions were obtained using atomic
structure codes and the Distorted Wave (hereafter DW)
approximation, similar to other opacity models [6-10]. 

 In addition to the limitations of ADOC work mentioned
above, new physical issues emerge in extending R-matrix calculations
towards a complete calculation of opacities. There are three major
problems that need to be solved: (A) convergence of large coupled channel
wavefunction expansions necessary to include sufficient atomic
structures manifest in opacity spectra, (B) completeness of high $n\ell$
contributions up to $n \equiv \infty$, and (C) attenuation of 
resonance profiles due to {\it intrinsic} autoionization broadening (included in
RM calculations in an ab initio manner) and {\it extrinsic} plasma
effects due to temperature and density, as generally considered for
bound-bound line opacity.
 
\subsection{Scientific problems}

 The erstwhile OP work summarized above concluded that the agreement
between OP and another independent calculation OPAL \cite{opal,op} 
do not differ by more than 2.5\%, implying that a further revision of
opacities was not needed \cite{b05}.
 However, there are outstanding problems related to opacities derived from the
 OP and all other opacity models. The foremost among them is related to a
downward revision of solar abundances of common volatile elements such
as carbon, nitrogen, oxygen and neon, relative to earlier ones by
up to $\sim$50\% \cite{agss09,aag21}. Thereupon, astrophysicists suggested that an {\it
upward} revision of opacities by $\sim$10\% \cite{bah05,sb04} would
countenance the lower solar abundances, since abundances are
inversely linked to opacities which affect the radiation field in
non-local thermodynamic equilibrium (NLTE) models employed to analyze
observed line profiles of elements. 
In particular, the iron opacity plays a crucial
role owing to relatively high abundance of iron.  
 Also, recent experimental measurements of iron opacity were higher than
given by OP and other models \cite{b15,n19}. Whereas opacity models have been
improved by including additional transition arrays
resonances, etc., the discrepancies with astrophysical and experimental
results remain outstanding.  

This series describes the work carried out since the 
OP opacities reported in 2003 and available via database OPServer
\cite{m07}.

\subsection{BPRM and DW Methods}
Current opacity models employ the DW approximation or
variants thereof. In order
to compare and contrast the present BPRM results, 
as well as to test complementarity and completeness 
of atomic data, we have also carried out relativistic distorted wave
calculations reported in paper RMOP3 of this RMOP series.
In principle, the DW approximation based on an atomic structure
calculation coupled to the continuum yields complete sets of
opacities. Oscillator strengths and photoionization cross sections are
computed for all possible bound-bound and bound-free transitions 
among levels specified by electronic configurations included in the
atomic calculation. However, since the DW approximation includes only 
the coupling
between initial and final states, the complexity of interference
between the bound and
continuum wavefunction expansions involving other levels is neglected.
That manifests itself as quasi-bound levels and autoionizing
resonances embedded in the continua. DW models employ
the independent resonance approximation that treats the bound-bound
transition probability independently from coupling to the continuum.
Apart from relative simplicity of atomic computations, 
the advantages of DW models is that
well-established line broadening treatments may be employed to
account for plasma interactions. Another advantage is ease of
completeness of datasets that can be augmented by including
additional configurations with multiple-electron excitations.
Furthermore, high angular-spin momenta do not pose a computational
problem commonly encountered in CC calculations. For these reasons the DW
method is generally employed for opacities calculations.
 In contrast, RMOP calculations are computationally laborious and 
time-consuming. However, coupling effects can affect atomic parameters
significantly. 

\subsection{Prior work}
 Opacity in the bound-free continuum is dominated by autoionizing
resonances, as shown in recently completed works cited above and present
results. Hitherto, they have been treated generally as lines
akin to
bound-bound transitions. The most important consequence, and likely
source of missing opacity, is the {\it intrinsic} autoionizing
broadening and the {\it extrinsic} plasma broadening thereof. The
much wider spread of resonances in the continuum than lines raises the
opacity significantly \cite{np16,p18}.

 Recent work \cite{d21} extended \fexvii R-matrix
calculations by including more configurations than NP16a.
Whereas that confirmed
our earlier results for photoionization cross sections, there are 
several issues: (i) D21 do not consider plasma broadening 
of autoionizing resonances that enhance opacities significantly,
(see papers II and III), (ii)
the D21 comparison between DW and unbroadened RM appears to agree, 
although fundamentally different since
the DW method treats autoionizing levels and broadening thereof as for
lines, (iii) D21 do not compare unbroadened RM cross sections
for \fexvii previously available from database NORAD \cite{norad}, (iv) 
inexplicably, D21 RM \fexvii Rosseland mean opacities are 
10\% below below the primarily DW results from OP2005, whereas 
all other DW models yield values up to 1.5 times higher \cite{p18};
there is no reason why
RM opacities, even without broadening, should be lower than OP and
other DW models, except that D21 might have an incomplete number of initial
\fexvii levels in their RM calculations. Other issues
such as radiative data, cross sections, and shapes of
autoionizing resonances due to plasma broadening 
are addressed in this series.

 Experimental opacity measurements at the Sandia Z facility
for Fe, Ni, and Cr
have highlighted deficiencies in theoretical models \cite{b15,n19}.
However,
experimental results need to be viewed in the context of the {\it very
limited
energy range where monochromatic iron opacity is actually measured}.
Indeed,the experimental energy range does {\it not}
include the region of maximum opacity from Fe ions around $\sim$ 1
Kev (well-known in X-ray spectroscopy). Therefore, experimental
opacities {\it per se} contribute only about 20\% to the
Rosseland mean opacities directly.
However, extrapolating the
differences between OP and experimental data in that limited range, B15
estimate a solar mixture opacity enhancement of 7$\pm$3\%
(the large error bars imply a factor of
2.5 discrepancy between the low and high experimental values).

  Opacity is a sensitive function of temperature and density, and an
incomplete tabulation in a limited range may give inconsistent results
since different ionization states of Cr, Fe and Ni contribute. For
example, N-like Cr ions with 3 active p-electrons make the largest
contribution at the Z temperature/density, whereas F-like Fe with 5
p-electrons is the largest contributor; for Ni it is Ne-like, a closed
p-shell configuration. These issues need to be examined individually at
a much wider range of energy-temperature-density to ascertain the source
of discrepancies.
Thus, although experimental results might point to "missing
physics", it is first important to include physics that is
known but missing, such as plasma broadening described in this series of
papers.

\subsection{Overview of RMOP calculations} 
Sections of this first paper P1 cover the
following topics, as well as general features of
subsequent papers in the series: 
(i) opacities and solar temperature-density structure,
(ii)
local-thermodynamic-equilibrium (LTE) plasma equation-of-state valid in
stellar interiors, (iii) relativistic effects using the
Breit-Pauli R-Matrix (BPRM) approximations, (iv) DW and BPRM calculations,
and (v) plasma broadening of autoionizing resonances in bound-free
opacity, (vi) convergence and completeness of atomic data.

\section{Monochromatic and mean opacities}

The atomic parameters comprising the monochromatic opacity are due to
 bound-bound (bb), bound-free (bf), free-free (ff), and
photon scattering (sc) contributions:
\begin{equation}
 \kappa_{ijk}(\nu) = \sum_k a_k \sum_j x_j \sum_{i,i'}
[\kappa_{bb(}(i,i';\nu) +
\kappa_{bf}(i,\epsilon i';\nu) + \kappa_{ff} (\epsilon i, \epsilon' i';
\nu) + \kappa_{sc} (\nu)]\ ,
\label{eq:k}
\end{equation}
where $a_k$ is the abundance of element $k$, $x_j$ the $j$ ionization
fraction, $i$ and $i'$ are the initial bound and final bound/continuum
states of the atomic species, and $\epsilon$ represents the electron
energy in the continuum. The atomic absorption coefficients are related
to the local radiation field at temperature T described by the Planck
function 

\be B_\nu(T) = \frac{(2h\nu^3/c^2)}{e^{h\nu/kT}-1}. \ee

 Macroscopic quantities such as radiative forces and fluxes
may be computed in terms of mean opacities, such as the Planck Mean
Opacity (PMO)

\be \kappa_P B(T) = \int \kappa_\nu B_\nu d\nu. \ee

 Of particular interest to opacity calculations is the Rosseland Mean
Opacity (RMO), $\kappa_R$ RMO defined as the {\it harmonic mean} of
monochromatic opacity $\kappa_{ijk}(\nu)$ as

\begin{equation}
 \frac {1}{\kappa_R} = \frac{\int_0^\infty g(u) \kappa_\nu^{-1}
du}{\int_0^\infty g(u) du} \ \ \ \hbox{\rm ; }\ \ \ g(u) = u^4 e^{-u}
(1 - e^{-u})^{-2},
\label{eq:RMO}
\end{equation}

where $g(u) = dB_\nu/dT$ is the derivative of the Planck weighting function
(corrected for stimulated emission). Eq.~\ref{eq:RMO} is mathematically
and physically a complex quantity to evaluate. Whereas the opacity 
determines radiative transfer
through the stellar interior, the RMO
is related to the total 
radiation flux that eventually escapes the star and observed \cite{nb2}. 
Although the 
singularity in the denominator $1/\kappa_nu$ is generally avoided owing
to overlapping spectral features, the RMO depends critically on the
precise distribution of monochromatic opacity at all frequencies at a
given (T,$\rho$) at each point inside the star. The opacity spectrum is
a complex quantity with superimposed dips or windows and large peaks
that vary by orders of magnitude due to energy dependence of atomic
parameters, 
$\kappa_{bb}(i,i') = (\pi e^2/m_ec)
N_i f_{ii'} \phi_\nu$, and $\kappa_{bf} = N_i \sigma_\nu$. The
$\kappa_\nu$ is then primarily a function of the bb oscillator strengths
$f$, bf  photoionization cross sections $\sigma_\nu$, level populations
$N_i$, and the line-profile factor $\phi_\nu$. The RMOP framework for
large-scale computations comprises mainly the first two components of
the opacity in Eq.~(\ref{eq:k}): (i) the bb transition probabilities and
(ii) the bf photoionization cross sections. 

\subsection{Solar structure and opacity}

 Tables 1 and 2 provide a numerical glimpse of solar interior structure and 
related plasma and atomic parameters. In Table 1 we focus on the region
outside of the nuclear fusion core in the
radiative zone up to the boundary at the base of the convection zone
(BCZ) \cite{nb1}. Helioseismological analysis of thousands of modes of
solar oscillations yields a precise measurement of the BCZ at 
solar radius R$_\odot$ = 0.713$\pm$0.001. At and above the BCZ outward
energy transport via radiative diffusion gives way to convection which 
becomes more efficient since $(dT/dr)_{diff} > (dT/dr)_{ad}$, the
adiabatic temperature gradient. There are
two main reasons for convective motions to be more efficient at the BCZ:
the weight of the
outer layers is less than the radiation pressure from the interior below, and
the {\it increase} in opacity from higher to lower temperatures. 
Opacity increases due to the prevalence of 
lower stages of ionization, as more bound electrons are active in 
absorption of radiation
via larger number of bound-bound and bound-free transitions than at
higher temperatures below the BCZ.

Table~2 shows the ionization states of the dominant elements that determine
opacity at the BCZ: O, Ne and Fe. Almost 90\% oxygen is in H-like 
or fully ionized, and 86\% of neon is in H-like and He-like 
ionization states. But one and two electron K-shell ionization states
do not contribute as much to opacity, as lower ones such as the partially
filled L-shell Fe ions with percentage contributions given in Table~2. 
Just three Fe ions constitute 85\% of iron at BCZ temperatures and densities.
Those ions, Fe~XVII, XVIII, XIX have very complex atomic structure and large 
number of radiative transitions that need to be accounted for. Large-scale
calculations are necessary to compute accurate opacities, and 
detailed calculations for these three ions are reported in RMOP2.

In table~3 we present a sample of the lowest and highest levels for the
\fexviii that has the highest ionization fraction of all Fe ions at BCZ
conditions (table~2). As described in paper RMOP2, RMOP calculations for
\fexviii yield 1,174 bound levels, and a total of 1,604 levels including 
high-lying levels with \en $>$ 4 described in paper RMOP4 
calculations to test
convergence and completeness. The MHD-EOS parameters
given in table~3 demonstrate 
the typical distribution of occupation probabilities and level
populations across the bound-level spectrum of complex Fe ions.
 Very high-lying levels make insignificant
contribution to opacity calculations. Previous works have discussed
the inexplicable differences of orders of magnitude in occupation
probabilities between OP and OPAL \cite{bs03}. The EOS issue therefore
remains open for future study. 

\begin{table}
\caption{Solar opacity parameters derived from \cite{b22} using
elemental abundances from \cite{agss09,aag21}
(numbers in parenthesis are powers of 10), with the exception of the
central temperature at $r/R_\odot$ = 0 from several other sources. The boundary of the radiative zone and base of the convection
zone(BCZ) is accurately determined from helioseismology to be 0.713$\pm$0.001
\cite{cd91,ba97}.}
\begin{center}
\begin{tabular} {|c|c|c|c|}
\hline
$r/R_\odot$ & $\rho$(g/cc) & T(K) & $N_e (cm^{-3})$\\
\hline
 0.00 & 162.2 & 1.58(7) & 1.0(26)\\
 0.35 & 6.89 & 5.75(6) & 3.57(24)\\  
 0.40 & 3.88 & 5.01(6) & 2.01(24)\\
 0.45 & 2.29 & 4.47(6) & 1.19(24)\\
 0.50 & 1.31 & 3.89(6) & 6.82(23)\\
 0.55 & 0.82 & 3.47(6) & 4.25(23)\\
 0.60 & 0.51 & 3.09(6) & 2.67(23)\\
 0.65 & 0.33 & 2.69(6) & 1.71(23)\\
 0.71 & 0.20 & 2.24(6) & 1.02(23)\\
\hline
\end{tabular}
\end{center}
\end{table}

\begin{table}
\caption{Main solar BCZ atomic opacity contributing elements and
ionization states and fractions $>$0.03, at T = $2.24 \times 10^{6}$K and
$N_e = 10^{23} cm^{-3}$, obtained from \cite{opcd} using the Q-form of
the MHD-EOS \cite{mhd,qmhd}. The actual elemental opacity contributions depend
significantly on the theoretical model employed
with respect to solar abundances and the EOS \cite{nb3}.}  
\begin{center}
\begin{tabular} {|c|c|}
\hline
 Element & Ionization state (fraction)\\
\hline
 Oxygen & O~VII (0.11), \  O~VIII (0.47), \ O~IX (0.42)\\
 Neon & Ne~VIII (0.10), \ Ne~IX (0.51), \ Ne X (0.35)\\
 Iron & Fe~XVI (0.031), \ Fe~XVII (0.196), \ Fe~XVIII (0.372), \ Fe~XIX
(0.284), \ Fe~XX (0.098)\\ 
\hline
\end{tabular}
\end{center}
\end{table}

\subsection{LTE equation-of-state}
 Stellar interiors are generally assumed to be characterized by a local
temperature-density (TD) parameter in LTE at any given point in the star.
However, TD tracks vary by orders of magnitude as nuclear energy
produced in the core is transported through the radiative diffusion zone
and the materially convective zones up to the atmosphere where radiation
escapes. A realistic EOS must therefore account for atomic-plasma effects
all throughout. In the first paper on OP opacities (\cite{symp}, hereafter SYMP),
the authors defined "stellar envelopes to be regions where atoms are
not markedly perturbed by the plasma environment"; the stellar envelope
generally comprising of radiative and convection zones.

   The MHD-EOS is a modified version of the
Saha-Boltzmann equations, based on the concept of {\it
   occupation probability} $w$ of an atomic level being populated,
   taking into account perturbations of energy levels by the plasma
   environment, 

 \be N_{ij} = \frac{N_j  g_{ij} w_{ij} e^{-E_{ij}/kT}}{U_j}. \ee

  The $w_{ij}$ are the occupation probabilities of levels $i$ in
  ionization state $j$. The occupation probabilities do not have a sharp
  cut-off, but approach zero for
  high-\en as they are "dissolved" due to plasma interactions.
 The partition function is re-defined as

 \be U_j = \sum_i g_{ij} w_{ij} e^{(-E_{ij}/kT)}. \ee

$E_{ij}$ is the excitation energy of level $i$, $g_{ij}$ its statistical
weight, and $T$ the temperature. The $w_{ij}$ are determined upon 
free-energy minimization in the plasma at a given temperature-density.
An atomic level $i$ is considered dissolved by the plasma microfield
when its highest Stark sub-level overlaps with the lowest sub-level of the
$i+1$ level (discussed further in RMOP3). 

 The original version of MHD-EOS estimated the range of validity to
$\rho <$ )0.02 g/cc. That rather restrictive density limit is less than
prevalent at the BCZ (c.f. Table~1), and most of the solar interior. The later
version called the Q-MHD \cite{qmhd} has been employed in all present 
calculations. However, the EOS employed by OPAL differs considerably
from OP; these differences and approximations made in the OP work,
have been previously discussed,
particularly for H-like ions for which data have been available
\cite{op,bs03}. Nevertheless, the agreement between OP and OPAL to $<5$\%
\cite{sb04} seems to indicate that the differences in EOS do not affect
final results. But the EOS redistributes level populations
significantly; further work on improving the MHD-EOS is in progress.

\begin{table}
\caption{MHD Equation-of-state parameters for \fexviii at solar BCZ: $T = 2
\times 10^6K, \ N_e = 10^{23}/cc$. Out of 1604 bound levels calculated,
the lowest six levels, energies, occupation probabilities W(OP), and 
percentage level populations are given. The highest bound levels
approaching the first ionization threshold E \ra 0 are also given. The rapid
decrease in W and N(\%pop) by orders of magnitude is evident.
Notation: 4.06(-5) = 4.06  $\times 10^{-5}$.} 
\begin{center}
\begin{tabular} {|c|c|c|c|}
\hline
 Level & Energy (Ry) & W(OP) & N(\% pop)\\
\hline
 $1s^22s^22p^5 (^2P^o_{3/2})$ & -99.924 & 1.00 & 8.79\\
 $1s^22s^22p^5 (^2P^o_{1/2})$ & -99.010 & 1.00 & 4.09\\  
 $1s^22s^22s(^1S_0) $  & -90.156 & 1.00 & 2.03\\
                       &         &      &      \\
 $1s^22s^22p^43s (^4P_{5/2})$ & -43.203 & 0.99 & 0.15\\
 $1s^22s^22p^43s (^4P_{3/2})$ & -42.957 & 0.99 & 0.05\\
 $1s^22s^22p^43s (^4P_{1/2})$ & -42.477 & 0.99 & 0.05\\
                       &         &      &      \\
Highest levels $n>4$ & -0.500    & 0.56 & 4.06(-5)\\
  Non-hydrogenic     & -0.343    & 0.01 & 1.10(-7)\\
\hline
\end{tabular}
\end{center}
\end{table}

\section{R-matrix opacity calculations}

 In this section we describe in some detail the differences 
from the OP work mentioned previously. In addition, a
description of the revised RMOP codes, related extensions, and the new set of
opacity codes is described. 

\subsection{Convergent close coupling calculations}
Owing to the fact that the OP R-matrix calculations 
included only a few
LS terms of the target or core
ion, the \eion wavefunction expansions were
far from convergence of computed
quantities and completeness. In the present work, particularly for iron
ions reported in RMOP2, an effort is made to ensure convergence of
photoionization calculations in the close
coupling (CC) approximation using the R-matrix method as developed in
the OP \cite{op}, and later in the Iron Project (IP) \cite{ip}.
In the CC approximation,
the atomic system is represented as the 'target' or the 'core' ion of
N-electrons interacting with the (N+1)$^{th}$ electron. The (N+1)$^{th}$
electron may be bound in the electron-ion system, or in the electron-ion
continuum depending on its energy to be negative or positive. The total
wavefunction, $\Psi_E$, of the (N+1)-electron system in a symmetry
$SL\pi$ or $J\pi$ is an expansion over the eigenfunctions of the target
ion, $\chi_{i}$ in specific state $S_iL_i(J_i)\pi_i$, coupled with the
(N+1)$^{th}$ electron function, $\theta_{i}$:
\begin{equation}
\Psi_E(e+ion) = A \sum_{i} \chi_{i}(ion)\theta_{i} + \sum_{j} c_{j}
\Phi_{j},
\end{equation}
where the sum is over the ground and excited states of the target or the
core ion. The (N+1)$^{th}$ electron with kinetic energy $k_{i}^{2}$
corresponds to a channel labeled
$S_iL_i(J_i)\pi_ik_{i}^{2}\ell_i(SL(J)\pi)$.
The $\Phi_j$s are bound channel functions of the (N+1)-electron system
that account for short range correlation not considered in the first
term and the orthogonality between the continuum and the bound electron
orbitals of the target.

Substitution of $\Psi_E(e+ion)$ in the Schrodinger equation
\begin{equation}
H_{N+1}\mit\Psi_E = E\mit\Psi_E
\end{equation}
introduces a set of coupled equations that are solved using the R-matrix
method. The solution is a continuum wavefunction $\Psi_F$ for an
electron
with positive energies (E $>$ 0), or a bound state $\Psi_B$ at a
{\it negative} total energy (E $\leq$ 0). The complex resonance
structures
in photoionization cross sections result from channel couplings between
the continuum channels that are open ($k_i^2~>$ 0), and ones that are
closed ($k_i^2~<$ 0). Resonances occur at electron energies $k_i^2$
corresponding to autoionizing states belonging to Rydberg series,
$S_iL_i\pi_i\nu \ell$ where $\nu$ is the effective quantum number,
converging on to the target threshold $S_iL_I$.

 Convergence of the \eion expansion in Eq.~1 is a difficult
computational problem in CC calculations, since the numerical size 
of the Hamiltonian
increases as square of the total number of channels in both the first
and the second sum on the RHS. For example, the calculations reported in
RMOP2 there are hundreds of target levels for each iron ion and thousands of
corresponding channels. 

\subsection{Relativistic effects and BPRM codes}

 The limited OP R-matrix calculations did not consider fine structure. 
However, subsequent IP work employed the BPRM 
framework \cite{ip,ben} including fine structure target levels and
recoupling scheme $LS \rightarrow LSJ$.
The relativistic BPRM Hamiltonian is given by
\begin{equation}
H_{N+1}^{\rm BP}= \sum_{i=1}\sp{N+1}\left\{-\nabla_i\sp 2 -
\frac{2Z}{r_i}
+ \sum_{j>i}\sp{N+1} \frac{2}{r_{ij}}\right\}+H_{N+1}^{\rm mass} + 
H_{N+1}^{\rm Dar} + H_{N+1}^{\rm so}.
\end{equation}
where the last three terms are relativistic corrections:
\begin{equation} 
\begin{array}{l}
{\rm the~mass~correction~term},~H^{\rm mass} = 
-{\alpha^2\over 4}\sum_i{p_i^4},\\
{\rm the~Darwin~term},~H^{\rm Dar} = {Z\alpha^2 \over
4}\sum_i{\nabla^2({1
\over r_i})}, \\
{\rm the~spin-orbit~interaction~term},~H^{\rm so}= Z\alpha^2 
\sum_i{1\over r_i^3} {\bf l_i.s_i},
\end{array} 
\end{equation}
respectively.

\begin{figure}
\centering{
\includegraphics[height=12.5cm,width=15cm]{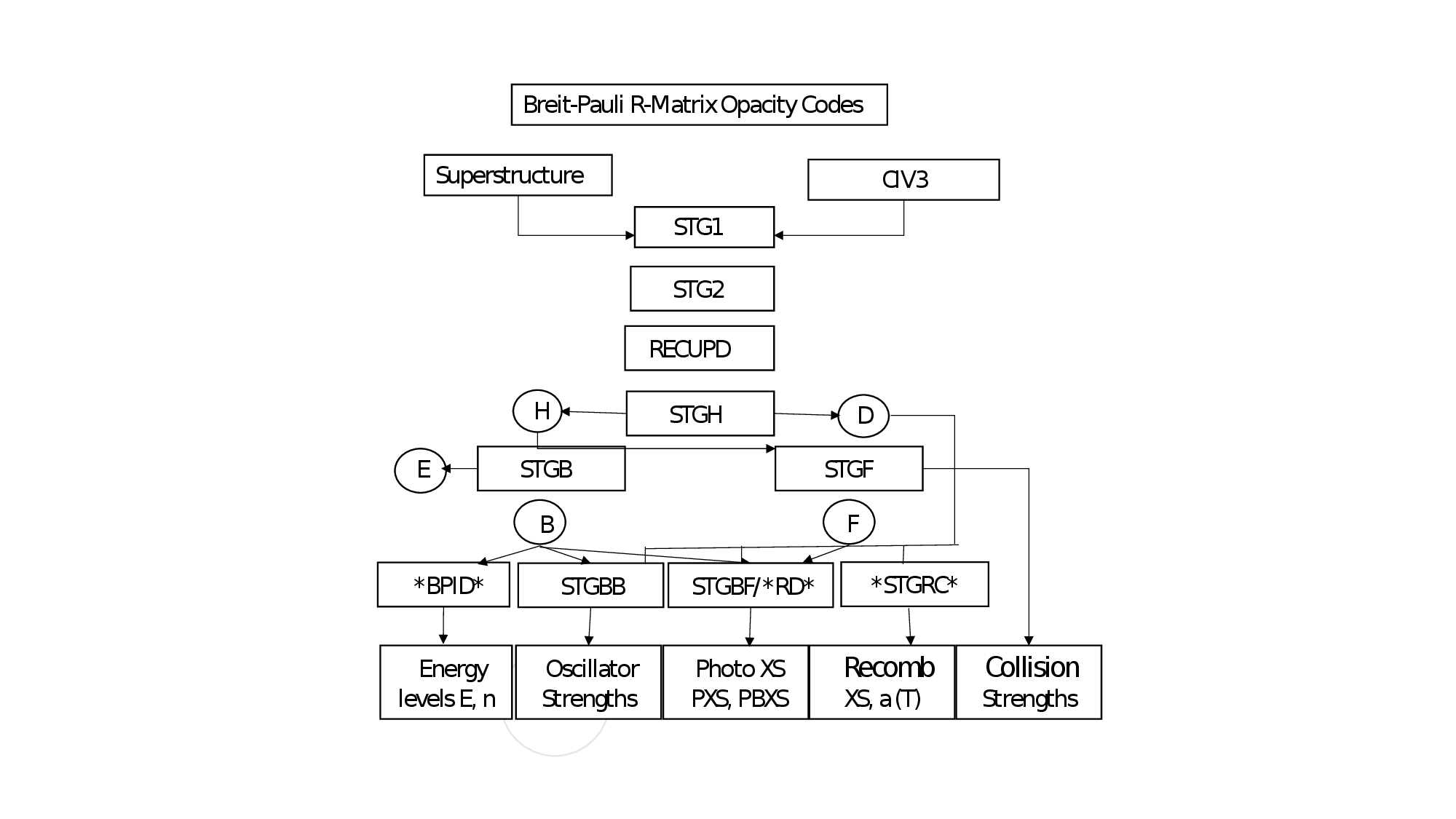}}
\caption{R-matrix codes for opacity calculations.
\label{fig:codes1}}
\end{figure}

 The BPRM codes used for the present opacity calculations 
are shown in Fig.~\ref{fig:codes1} which is modified from the LS
coupling version given in ADOC~II \cite{adoc2}. The atomic structure
codes Superstructure (SS), CIV3, STG1, STG2, STGH, STGB, STGF, STGBB and
STGBF are described in \cite{adoc2}. Briefly, SS and CIV3 are atomic
structure codes; either one is first employed to obtain reasonably accurate
target wavefunctions, eigenenergies, and oscillator strengths for the
target ion. The target ion orbital radial functions are then used by STG1
to reconstruct the target and 
calculate R-matrix basis functions and radial integrals for the
\eion system. With radial integrals from STG1 as input, 
STG2 computes angular coefficients and matrices of the Hamiltonian and
dipole operators. 
The BP recoupling $LS \rightarrow LSJ$ is implemented in the code
RECUPD \cite{ben}. STGH diagonalizes the BP Hamilitonian and produces
the H and D files required to obtain physical parameters such as energy
levels and radiative data such as oscillator strengths and 
photoionization cross sections. 

Other new codes or extended
versions (bracketed by asterisks) comprise of the following. 

\subsection{Level identification} Energy levels from R-matrix
calculations in STGB are obtained as eigenvalues of bound states without
spectral designation, as in atomic structure calculations. 
The new code BPID is employed to assign spectroscopic identification of
all computed fine structure levels. Following
diagnolization of the Hamiltonian matrix, the R-matrix basis functions
are obtained and used to compute energy levels in STGB. 
BPID then analyzes the parameters computed in STGB to determine
spectroscopic identification. Those are the 
channel percentage weights and quantum defects, complemented independently
by atomic structure calculations. Level identification is necessary not
only for spectroscopic designations required in practical applications, 
but also for matching and high $n\ell(SLJ)$ 
"top-up" of computed oscillator
strengths from STGBB and photoionization cross sections from STGBF to
test completeness of atomic data. 

\subsection{Radiation damping} For highly charged ions, in particular
H-like and He-like ions of Fe-group elements, radiative damping of
autoionizing resonances is important (e.g. \cite{pz97}), and may
considered using the extended code STGBF-RD. However, for opacities
calculations this is not needed since total photon absorption cross
section regardless of subsequent radiative decays is required.
 
\subsection{Unified \eion recombination} Level-specific and total
\eion recombination cross sections may be computed employing the unified
method
subsuming both radiative and di-electronic recombination in {\it
ab initio} manner within
the R-matrix CC formulation \cite{np92}, 
using the code STGRC (\cite{aas}, and
references therein).

\subsection{Convergence and Completeness}

Even when the CC calculations are converged to practically acceptable
accuracy, as discussed in RMOP2, completeness may not have been achieved
with respect to all of the bound-bound and bound-free transitions in an
ion.
 But R-matrix calculations become computationally intensive with increasing
energy as successive thresholds of the target ion are exceeded and more
channels open up. At the same time 
computations need to be done at all energies with 
a sufficiently fine energy mesh to resolve autoionzation resonance
structures. However, above the highest target level all channels are 
open and there are no more resonances. Although the number of open
channels may be large, the cross sections are featureless and slowly
varying with energy. Moreover, for high $n\ell(SLJ)$ levels resonance
structure are weak and may be neglected. In such cases a "top-up"
procedure using DW methods may be employed to test if convergence has
been achieved to ensure completeness of atomic data for opacities. One
such "top-up" procedure is described in paper RMOP4. Generally, we find that
the "top-up" contribution to opacities is small and does not exceed
$\sim$5\% for any given ion.

\subsection{Plasma effects}
The OP and RMOP calculations are carried out for isolated atomic
systems. As such, external effects due to plasma environment at specific
temperature, density, abundances, etc. need to be considered in opacity
calculations. Those effects determine the EOS as discussed above.
But in addition, they alter computed atomic features such as line shapes
of bound-bound atomic transitions significantly. 

 In OP work quasi-bound levels that give rise to resonances in the
continuum are treated as bound levels {\it a priori},
and plasma broadening of autoionizing resonances is neglected due
to, (i) difficulty in including pressure broadening, and (ii) because
quantum interference between resonances and the continuum is considered
to be small
\cite{bs03}. Therefore, a perturbative approach in the independent
resonance approximation, akin to independent treatment of radiative and
di-electronic recombination, is employed. However, as we demonstrate in
paper RMOP3 in detail, plasma broadening of 
autoionizing resonances fundamentally different from that of line 
broadening of bound levels. Practically, plasma broadening
not only has a significant but large effect on bound-free opacity and
derivative quantities such as the Rosseland and Planck mean opacities.

\subsection{Opacities calculations}

\begin{figure}
\centering{
\includegraphics[height=12.5cm,width=15cm]{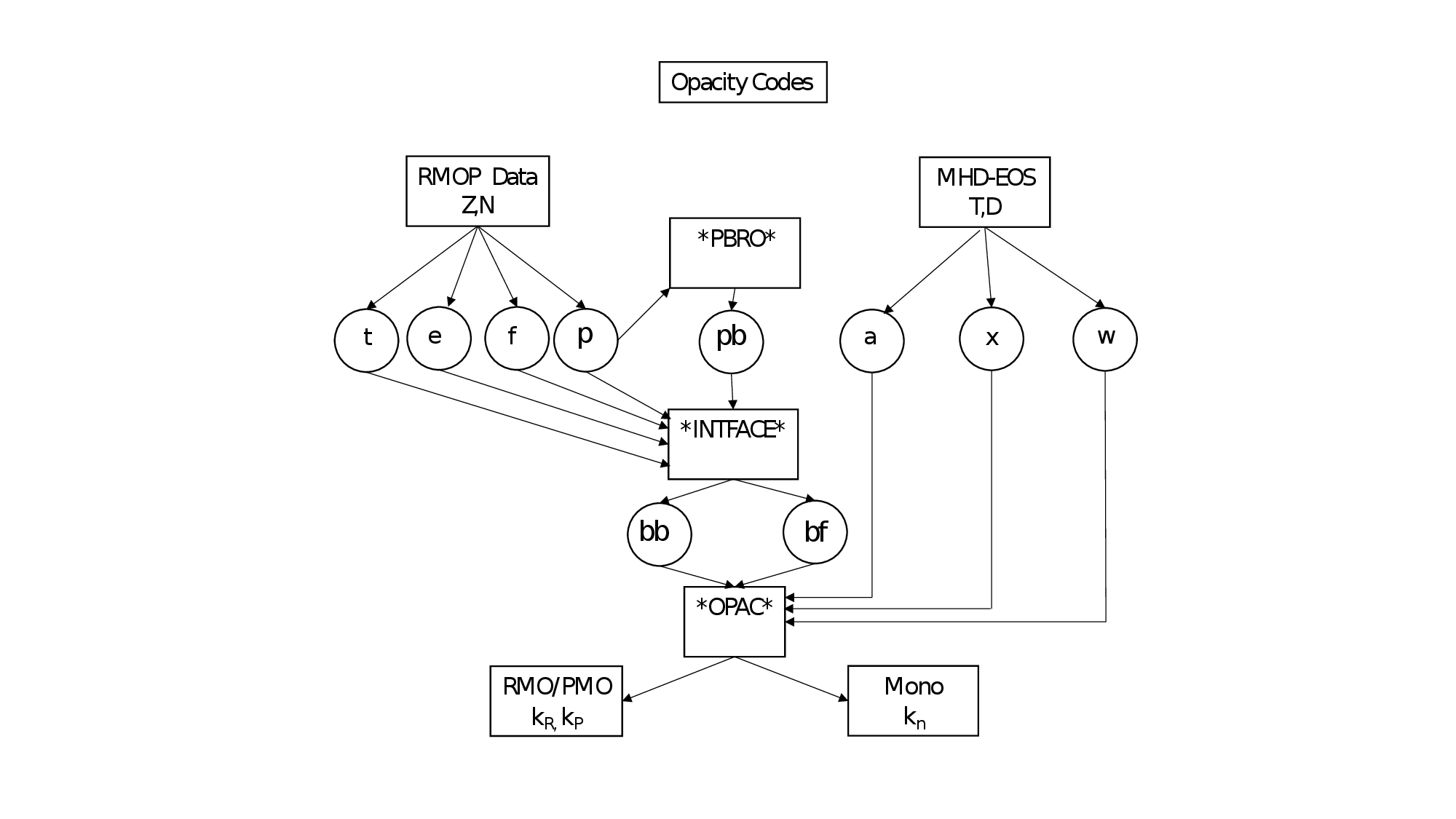}}
\caption{Plasma opacity codes.
\label{fig:codes2}}
\end{figure}

The opacity codes employed in RMOP calculations have not heretofore been
published, and are different from those
in the OP work. In the initial stages of OP, both sets of codes have been 
extensively checked against each other for opacities reported in
\cite{symp}. However, most of OP data was from
sources other than R-matrix calculations and processed to compute
opacities in a different manner than described herein.
Fig.~\ref{fig:codes2} shows the schematic diagram of the codes and
datasets in RMOP calculations (codes bracketed by '$\ast$' have not been
heretofore presented).

\subsubsection{Atomic data}
 The input RMOP data for opacity calculations
are the final products from codes shown in Fig.~\ref{fig:codes1}.  
Each ion is treated
as an \eion system
characterized by (Z,N), the atomic number Z and the number of electrons
in the target ion.

The input atomic datasets consists of four
files: (i) t-file --- target level energies and statistical weights, (ii)
e-file --- energy
levels as computed by STGB and further processes using BPID, (iii)
f-file --- oscillator strengths for E1 transitions computed in STGBB,
(iv) p-file --- photoionization cross sections from STGBF. These files
are input to the code INTFACE that interfaces the atomic data, and  maps out
at a photon frequency mesh of 100,000 frequencies (in contrast the OP
work is at 10,000 frequencies), into bound-bound (bb) and bound-free (bf) 
files for opacity calculations separately for each ion.
{\it Prior to input into INTFACE, the p-files are pre-processed by the code
PBRO for plasma broadening of autoionizing resonances in photoionization
cross sections to produce broadened bf-files.}
PBRO computes plasma broadened cross sections (described in RMOP3) for each
temperature and density. This results in a large number of pb-files for
all TD pairs from
a single unbroadened p-file for
each level of each ion of each element in opacities calculations.
INTFACE then processes either the unbroadened p-files or broadened pb-files
and produces corresponding bf-files mapped on to the opacity frequency
mesh. Thus, a huge amount of data is produced as result of the
interface of atomic and plasma parameters, most of it too large to be
stored and therefore treated as intermediate files that are recreated
for each TD. 

\subsubsection{Equation-of-state}
The MHD-EOS parameters are taken from OPCD codes
using the Q-form \cite{mhd,qmhd,op}, so that there are no inconsistencies owing
to the EOS between RMOP and OP. 
The input EOS parameters consist of: a
--- abundances of elements, x --- ionization fractions of each ion, and
w --- EOS data to obtain occupation probabilities.
The opacity code OPAC computes monochromatic, Planck and Rosseland mean
opacities, $\kappa_\nu, \kappa_P, \kappa_R$ respectively,
 using the INTFACE bb and bf files, and EOS parameters,
independently for each T-D, or ranges thereof. Since one of the primary
motivation of the RMOP calculations is to solve the aforementioned
solar abundances problem, different sets of abundances may be used to
ascertain differences among them.

\subsubsection{Bound and continuum opacities}
  The most important difference between RMOP and other opacity
calculations is the treatment of bound-bound as distinct from
bound-free continuum opacity. There is a clear division between lines as
strictly the transitions among negative energy bound levels, and
autoionizing resonances in the bound-free continua. Practically, this
difference manifests itself in the code
OPAC (Fig.~2). The bb-opacity consists of negative energy bound levels 
only, and corresponding oscillator strengths, and the bound-free opacity
consists of photoionization cross sections with resonances that are
otherwise treated as lines in DW opacity calculations. The DW
calculations may couple lines {\it a posteori} 
to single-channel feature-less continuum perturbatively, but not in a
fully coupled manner as in RMOP opacities. The combined bb and bf opacity
spectra therefore are quite different in detail, which reflects in the
calculation of mean opacities.

 There are additional steps necessary in order to ensure completeness,
as discussed in RMOP4, relating to the division between negative and
positive energy levels and to ensure that there is no double-counting of
levels if it is necessary to include high-$n\ell$ contributions,
although they are found to matter little since high-lying levels have
insignificant populations.

 The treatment of free-free contribution to plasma broadening, discussed
in RMOP3, is also implemented in OPAC. 
Large datasets of $f$-values for transitions {\it among
positive energy levels}, obtained from atomic structure codes such as
Superstructure or variants, are required to compute this contribution.
Although, small relative to electron impact, Stark and Doppler broadening,
it nevertheless needs to be included for completeness.

\section{Acknowledgments} This work has been partially supported by
grants from the US National Science Foundation, NASA, and the Department
of Energy. Most of the computational work was carried out at the Ohio
Supercomputer Center.


\begin{thebibliography}{20}
\bibitem{nb3} Oxygen opacity stems mainly from fully ionized or H-like
O~VIII and therefore unlikely to change because of the accuracy of 
atomic data.
\end{thebibliography}
\end{document}